\documentstyle[aps,prl,floats]{revtex}
\input{psfig}
\begin{document}
\draft
\twocolumn[\hsize\textwidth\columnwidth\hsize\csname@twocolumnfalse\endcsname
\title{Current-Induced Step Bending Instability on Vicinal Surfaces }
\author{Da-Jiang Liu$^1$, John D. Weeks$^{1,2}$, and Daniel Kandel$^3$}
\address{$^1$Institute for Physical Science and Technology, University
of Maryland, College Park, Maryland 20742}
\address{$^2$Department of Chemistry,
University of Maryland, College Park, Maryland 20742}
\address{$^3$Department of Physics of Complex System,
Weizmann Institute of Science,
Rehovot 76100, Israel}
\date{\today }
\maketitle

\begin{abstract}
We model an apparent instability seen in recent experiments on current
induced step bunching on Si(111) surfaces using a generalized 2D BCF
model, where adatoms have a diffusion bias parallel to the step edges
and there is an attachment barrier at the step edge.  We find a new
linear instability with novel step patterns.  Monte Carlo simulations
on a solid-on-solid model are used to study the instability beyond the
linear regime.  

\end{abstract}

\pacs{68.35.Ja,68.10.Jy,68.55.Jk,05.70.Ln}
]

Vicinal crystal surfaces can exhibit a number of different
morphological instabilities that may be important in crystal growth
and device fabrication. One of the most interesting of these
instabilities arises from the biased diffusion of adatoms under the
influence of an external driving force, as illustrated by the step
bunching instability on Si(111) surfaces~\cite{latyshev} induced by
heating with a direct electric current. Many different aspects of this
so-called electromigration problem have been investigated when the
driving force is normal to the steps: e.g., the microscopic origin of
the driving force on adatoms, the effect of temperature on the
instability, and the resulting step
patterns~\cite{stoyanov,misbah,kandel95,kandel96}. We show here that
when the relative orientation of the external driving force and the
miscut angle of the vicinal surface is varied, there exists a
qualitatively different instability that can be explored both
experimentally and theoretically.

Because of the intrinsic anisotropy of a stepped surface, kinetic
instabilities can usually be associated with one of the two principal
axes normal to and along the average step orientation (the $x$ and $y$
directions, respectively). It is convenient to describe step positions
in Fourier space. From a linear stability analysis, the evolution of
small perturbations in the positions of the steps away from a uniform
step train $\delta x_n(y,t)$ can be written as
\begin{equation}
\delta x_n(y,t)=\sum_{q,\phi }\delta x(q,\phi )e^{iqy+in\phi +\omega (q,\phi
)t}+{\rm c.c.,}  \label{Fourier1}
\end{equation}
where $q$ is a wavenumber along the step direction, and $\phi $ a phase
factor relating consecutive steps. When ${\rm Re}\,\omega (q,\phi )>0$, a
perturbation of that mode will grow exponentially.

One class of instabilities involves the growth of finite $q$
perturbations along the $y$ (step)\ direction, as illustrated by the
Bales-Zangwill~\cite{bales} instability. In this example it is
sufficient to take $\phi =0$, since this (in-phase) mode is maximally
unstable and will dominate. Indeed the underlying (Mullins-Sekerka)
instability can be understood by considering a single interface.

Another class of instabilities involves the relative motion of
neighboring steps, and can usually be understood using a one
dimensional (1D)\ model with $q=0$ and the step pairing mode $\phi
=\pi .$ Examples include the step bunching instability due to either
the Ehrlich-Schwoebel barrier~\cite{ehrlich66,schwoebel}, impurity
adsorption~\cite{frank,eerden86,kandel95a}, or the usual
electromigration instability with the current normal to the
steps. Stoyanov~\cite{stoyanov} used a 1D model to study the latter
instability in detail. He showed that a vicinal surface can be
unstable towards step bunching when adatoms have a diffusion bias in
the step-down direction and there exists an extra barrier for
attachment to step edges.  The step pairing mode ($\phi =\pi $) gives
the maximum instability unless there are very strong repulsions
between steps~\cite{sato} or significant direct interterrace
diffusion~\cite{liu98,stoyanov2}.

Quantitative studies of electromigration based on a generalized 1D
step model have shown very good agreement with many aspects of the
experiments~\cite{liu98}. However, a more general perspective is
required to understand some recent surprising results by Latyshev {\em
et al.}~\cite{latyshev94}.  They observed the sudden appearance of
{\em step bends} in certain regions of the surface where the step
edges are nearly parallel to the electric current, leading to the
formation of {\em anti-step} (down-step) bunches as well as the usual
(up-) step bunches. Motivated by this result, we study here a 2D step
model with biased diffusion {\em parallel} to the step edge direction,
and indeed find an instability. This new instability requires {\em
both} finite $q$ and finite $\phi $ and reaches its maximum when the
phase shift $\phi =\pm \pi /2$ at small $q$, unlike any other surface
instability we are aware of.

\begin{figure}[tbp]
\centerline{\psfig{file=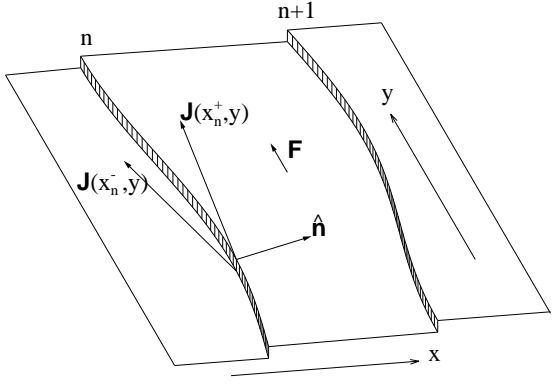,width=3in}}
\caption{Schematic view of a vicinal surface with an external force
parallel to the step edge.}
\label{notation}
\end{figure}

Figure \ref{notation} gives a schematic view of the surface and
introduces the notation. As in Stoyanov's model~\cite{stoyanov},
adatoms have a drift velocity $D_sF/kT$, where $D_s$ is the adatom
diffusion rate on flat terraces. $F$ is the magnitude of the external
force acting on adatoms, and we assume in this case it is in the $y$
direction. For simplicity, here we ignore evaporation and deposition
of adatoms on terraces; this does not affect the basic instability we
find. We make the usual quasistatic approximation~\cite{langer},
setting $\partial c/\partial t=0$ in the diffusion equation describing
the adatom concentration field $c(x,y)$ on the terraces:
\begin{equation}
D_s\left( \frac{\partial ^2c}{\partial x^2}+\frac{\partial
^2c}{\partial y^2}\right) -\frac{D_sF}{k\,T}\frac{\partial c}{\partial
y}=0.  \label{diffusion}
\end{equation}
The boundary conditions are given by linear kinetic expressions with a
parameter $\kappa $ governing the rate of exchange of atoms between
steps and terraces:
\begin{equation}
\pm \kappa [c_n^{{\rm eq}}(y)-c(x_n^{\pm }(y),y)]=\hat{{\bf n}}\cdot
{\bf J} (x_n^{\pm }(y),y), \label{boundary}
\end{equation}
where the surface adatom flux ${\bf J}$ is given by 
\begin{equation}
{\bf J}(x,y)=-D_s\left( -{\partial }/{\partial x},-{\partial
}/{\partial y} +f\right) c(x,y),
\end{equation}
and $f\equiv F/kT$. The $+(-)$ indicates the ascending (descending)
terrace next to step $n$ as shown in Fig.~\ref{notation}. $c_n^{{\rm
eq}}(y)$ is the local equilibrium adatom concentration of step $n$ at
position $y$, and $\hat{{\bf n}}$ is the unit vector normal to the
step edge. Through microscopic mass conservation, the normal velocity
of step $n$ can be written as
\begin{eqnarray}
v_n &=&a^2\hat{{\bf n}}\cdot \left[ {\bf J}(x_n^{+},y)-{\bf J}
(x_n^{-},y)\right] \nonumber \\ &=&D_sa^2[2\,c_n^{{\rm
eq}}(y)-c(x_n^{+},y)-c(x_n^{-},y)], \label{velocity}
\end{eqnarray}
where $a$ is the lattice constant in the $x$-$y$ plane, assuming a
square lattice.

To take into account the stabilizing effects of the line tension and
step repulsions, we write
\begin{equation}
c_n^{{\rm eq}}(y)=c_0^{{\rm eq}}\exp \left( \frac{\mu _n(y)-\mu
_C}{k\,T} \right) , \label{ceq}
\end{equation}
where the step edge chemical potential $\mu _n(y)$ --- the change in
free energy per atom for adding atoms to the step at position $y$ ---
is given by~\cite{liu97b,weeks97}
\begin{equation}
\mu _n=a^2\left[ V^{\prime }(w_n)-V^{\prime
}(w_{n-1})+\tilde{\beta}\partial ^2x_n/\partial y^2\right] +\mu _C.
\label{mu}
\end{equation}
Here we assume there is a nearest-neighbor repulsive interaction
$V(w)$ between steps separated by a terrace width $w$ and $V^{\prime
}(w)=dV/dw$. $\tilde{\beta}$ is the step edge stiffness and $\mu_C$ is
the equilibrium atom chemical potential of the crystal.

To carry out a linear stability analysis~\cite{linearnote}, we
substitute the solution of the diffusion equation
Eq.~(\ref{diffusion}) with boundary conditions Eq.~(\ref{boundary})
into Eq.~(\ref{Fourier1}). Keeping terms linear in $\delta x$, we find
for the growth rate $\omega(q,\phi)$:
\begin{equation}
\omega (q,\phi )=\frac{-2D_sc_{eq}^0a^2\Lambda _q\left[ f \,q\,d\,\sin
\phi +g(q,\phi )\right] }{2\Lambda _qd\cosh (\Lambda _q w_0)+\left(
\Lambda _q^2\,d^2+1\right) \sinh (\Lambda _qw_0)}, \label{omega}
\end{equation}
where 
\begin{eqnarray}
g(q,\phi ) &=&g_x(1-2\cos \phi +\cos 2\,\phi )-g_yq^2\cos \phi
\nonumber \\ &&+\left[ g_yq^2+2g_x(1-\cos \phi )\right] \nonumber \\
&&\times \left[ \cosh (\Lambda _qw_0)+\Lambda _qd\sinh (\Lambda
_qw_0)\right] ,
\end{eqnarray}
\begin{equation}
\Lambda_q = \sqrt{q^2 + i f q},
\end{equation}
and 
\begin{equation}
g_x=\frac{a^2}{k\,T} \frac{d^2V}{dw^2} \bigg\vert_{w=w_0}\,;\quad
g_y=\frac{\tilde{\beta}a^2}{k\,T}.
\end{equation}
In the above equations there is an important length scale $d\equiv
D_s/\kappa $, the ratio between the adatom diffusion rate and the
attachment rate, as introduced by Pimpinelli {\em et
al.}~\cite{pimpinelli}. $g(q,\phi ) $ is the stabilizing term from the
line tension and step repulsions; it reduces to the known result of
Pimpinelli {\it et al.}~\cite{pimpinelli} in the case when $g_x=0$
(i.e., ignoring step repulsions) and $f \rightarrow 0$.

\begin{figure}[tbp]
\centerline{\psfig{file=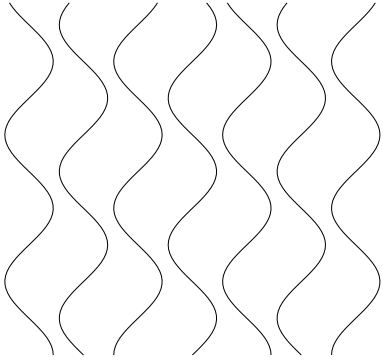,width=2in}}
\caption{Illustration of a step train with sinusoidal fluctuations
with phase shift $-\phi=\pi/2$. }
\label{phase}
\end{figure}

The first term in the square bracket of Eq.~(\ref{omega}) is the major
result of this paper. For $f >0$ (external force in the $+y$
direction), ${\rm Re} \, \omega$ is positive for $-\pi <\phi <0$ and
is maximally unstable when $\phi =-\pi /2$ for a particular wavenumber
$q$ that is not too large [for large enough $q$, the relaxation term
$g(q,\phi)$ will dominate]. This is quite different from the usual
step pairing instability with maximum instability at $\phi =\pi $ if
the bias is in the step down direction.  Figure \ref{phase} is an
illustration of a step train with $\phi=-\pi/2$.  For several
neighboring steps, the hills and troughs are shifted in their
positions and form regions of high step densities (narrow terraces)
that extend along certain direction. Changing the sign of $f$ (the
direction of the external force) makes modes with $0<\phi <\pi $
unstable. Note that if the attachment rate is very fast (so that
$\kappa \rightarrow \infty $ or $d\rightarrow 0$), the mass transport
is limited by the diffusion rate and the instability disappears. Thus
the extra barrier for adatom attachment to step edges is crucial for
the existence of the instability.

\begin{figure}[tbp]
\centerline{\psfig{file=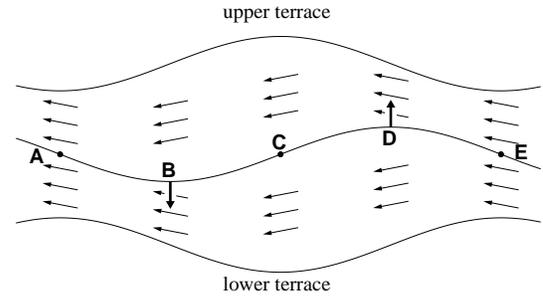,width=3in}}
\caption{Illustration of the atomic flux induced by an external field
parallel to the average step edge direction on two terraces bounded by
three steps with sinusoidal perturbations and phase shift $\phi =-\pi
/2$. }
\label{qual}
\end{figure}

We can gain a better understanding of the physical origin of this
instability from the following qualitative picture. Figure \ref{qual}
draws three steps with perturbations in the form of sinusoids with a
phase shift $\phi =-\pi /2$ (steps are out-of-phase with every other
step). We assume there is a very strong barrier for adatom attachment
to a step edge. The direction of the atomic flux on a terrace driven
by an external force parallel to the average step edge direction is
then approximately the average of the directions of the two steps
bounding the terrace.

From microscopic mass conservation [Eq.~(\ref{velocity})], the
velocity of a segment of a step is given by the difference of the
normal components of the two surface atomic fluxes at the two sides of
the step. For segments $A$, $C$, and $E$ in Fig.~\ref{qual}, the two
fluxes cancel and the velocity vanishes at those points. However for
segments $B$ and $C$, there is a net contribution from the two fluxes
that makes the step move in the direction shown. Thus the amplitude of
the perturbation will increase and the step train is
unstable. Changing the direction of the external force will reverse
the direction of the motion and stabilize the perturbation of
Fig.~\ref{qual}, but will destabilize a perturbation with phase
shift $\phi =\pi /2$. Thus a vicinal surface is unstable when there is
driven diffusion {\em parallel} to the step edge direction. This
picture is only correct when there are no efficient channels for
adatoms to hop directly from one terrace to another one.

It is natural and straightforward to extend the linear stability
analysis to an external force in a general direction. The general
solution is too complicated to write down here, but in the limit of
weak external forces and small wavenumber $q$, we have
\begin{equation}
\omega \approx \frac{2 D_s c^0_{{\rm eq}} a^2}{2 d + w_0} \left[- 2 f_x d
\frac{1 - \cos \phi}{2d + w_0} - f_y q \,d \sin \phi - g(q, \phi)
\right]
\label{fxy}
\end{equation}
where $f_{x(y)} = F_{x(y)}/kT$ is the $x(y)$ component of the
(reduced) external force. The term proportional to $f_x$ has the
familiar $(1- \cos \phi)$ form for the pairing instability for an
external force normal to the steps.

These results may help us understand the origin of the anti-step
bunches seen by Latyshev {\it et al}{\em .}~\cite{latyshev94} on
Si(111) at elevated temperatures. Initially the current direction is
{\em perpendicular} to the average step direction and step bunching
occurs in the usual way. Because of sublimation, nearly uniform arrays
of {\em crossing steps} form between the step
bunches~\cite{kandel95}. As the system coarsens and the distance
between bunches gets larger, the inclination angle of the crossing
steps increases and becomes nearly perpendicular to the step bunches,
and thus parallel to the current direction. It is at this stage that
Latyshev {\em et al.}~\cite{latyshev94} observe the appearance of {\em
step bends} in the crossing arrays that grow in size and accumulate
into an anti-step bunch, i.e., a step bunch in the opposite direction
to the initial orientation of the vicinal surface.

Since the physical model discussed in this paper also give good
agreement with the initial step bunching behavior of
Si(111)~\cite{liu98} when the current is normal to the steps, it seems
quite likely that the instability discussed above is responsible for
the appearance of the step bends in the crossing arrays. However, a
detailed analysis of the anti-step bunch formation in the experiments
necessarily involves the nonlinear regime and a treatment of
interactions with the preexisting step bunches, which is beyond the
scope of this letter.

\begin{figure}[tbp]
\vspace{-0.5cm}
\centerline{\psfig{file=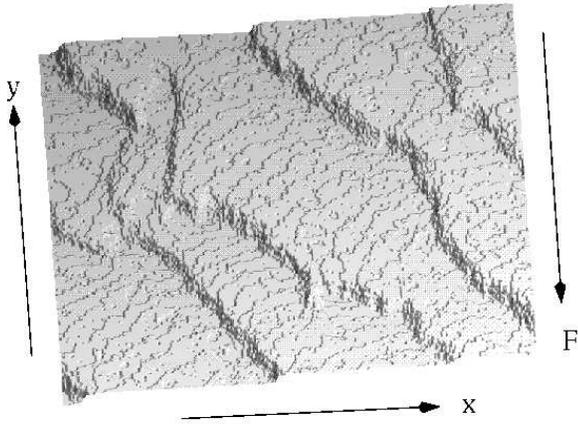,width=3.in}}
\caption{Snapshot of a simulation using a solid-on-solid model with a
diffusion bias parallel to the step edge direction. The initial steps
are along the vertical ($y$) direction. }
\label{sos16biasb2}
\end{figure}

To get some information about the nonlinear behavior, we carried out
kinetic Monte Carlo simulations of a solid-on-solid (SOS)
model~\cite{liu98} that incorporates the essential physical features
of biased diffusion with attachment barriers at steps. The external
force is mimicked by an asymmetry in the attempt frequencies for
exchange of surface atoms between nearest-neighbor sites. The energy
barrier is simply the binding energy of the surface atom in its
initial site, except that there is an {\em extra} barrier when the
movement causes a change in total energy, and hence for attachment at
steps (see Ref.~\onlinecite{liu98} for more
details). Figure~\ref{sos16biasb2} is a snapshot of a simulation with
the driving force parallel to the average step edge direction. The
initial step edges are in the vertical direction and adatoms have a
downward diffusion bias. The dark regions are step bunches formed by
step segments bending in the opposite direction to the individual
steps (black lines) on the terrace. Note that only a relatively small
region of each step lies within the step bunch, in contrast to the
usual step bunches that form when the current is normal to the step
direction. This general feature is already suggested from the patterns
in Fig. \ref{phase}.

The earlier SOS simulations of Dobbs and Krug~\cite{dobbs96} gave
very different results. They used Metropolis dynamics, which does not
incorporate the step edge barrier, and the resulting step patterns
contained several artifacts and did not compare well with experiments
on Si(111).

Krug and Dobbs~\cite{krug,dobbs96} also introduced a continuum model
that showed that a surface is quite generally unstable under an
external driving force when the adatom mobility has a slope
dependence.  This prediction, unlike other features, e.g., the
orientation of selected facets and the coarsening rate, is independent
of the underlying physical origin of the slope dependence (in our
case, the step edge barrier) and is thus robust.  Indeed one can make
connection between the step model and the continuum model by taking
$\phi \rightarrow q_x w_0$ and $q \rightarrow q_y$ and expanding
Eq.~(\ref{fxy}) for small $q_{x(y)}$:
\begin{equation}
\omega \sim f_x q_x^2 + (2 d + w_0) w_0^{-1} f_y q_x q_y + {\cal O}(q^4)
\end{equation}
Except for an anisotropic ratio that is absent from their model, this
equation is qualitatively the same as the result of the continuum
model in Ref.~\onlinecite{dobbs96} for a surface tilted away from a
high symmetry plane with the broken symmetry of the underlying crystal
structure along the $x$ direction.  Although their continuum model
does not capture all the essential physics of a vicinal surface below
the roughening temperature, e.g., the anisotropy in the adatom
mobility and the singularity in surface stiffness at zero slope, it
does give an idea of the instability at very large length scales.

In conclusion we report a new linear instability on vicinal surface
that arises from the interplay between a step train and the two
dimensional diffusion fields between the steps. There are many
interesting theoretical issues yet to be resolved. The study of the
instability beyond the linear regime is needed to understand the
coarsening behavior and anti-step bunch formation. The effect of a
current along the step direction is also being investigated
experimentally~\cite {latyshev98}.

We are grateful to H.-C. Jeong, O. Pierre-Louis, A. V. Latyshev, and
E. D. Williams for helpful discussions.  This work has been supported
by the National Science Foundation under Grant No. DMR-9632521 and by
grant No. 95-00268 from the United States-Israel Binational Science
Foundation (BSF), Jerusalem, Israel. DK is the incumbent of the Ruth
Epstein Recu Career Development Chair.

\end{document}